\documentclass[preprint,aps,superscriptaddress,byrevtex,notitlepage,endfloats*]{revtex4}
\usepackage{psfig}

\begin{document}

\title{Polymers near Metal Surfaces: Selective Adsorption and
Global Conformations}

\author{L. Delle Site}
\affiliation{Max-Planck-Institute for Polymer Research, P.O. Box 3148,
  D-55021 Mainz, Germany}
\author{C. F.Abrams}
\affiliation{Max-Planck-Institute for Polymer Research, P.O. Box 3148, D-55021 Mainz, Germany}
\author{A. Alavi}
\affiliation{Department of Chemistry, Cambridge University, Lensfield
  Road, Cambridge CB2 1EW, U.K.}
\author{K. Kremer}
\affiliation{Max-Planck-Institute for Polymer Research, P.O. Box 3148, D-55021 Mainz, Germany}

\date{\today}

%\maketitle

\begin{abstract}
We study the properties of a polycarbonate melt near a nickel surface
as a model system for the interaction of polymers with metal surfaces
by employing a multiscale modeling approach. For bulk properties a
suitably coarse grained bead spring model is simulated by molecular
dynamics (MD) methods with  model parameters directly derived from
quantum chemical calculations. The surface interactions are
parameterized and incorporated by extensive quantum mechanical density 
functional calculations using the Car-Parrinello method. We find
strong chemisorption of chain ends, resulting in significant
modifications of the melt composition when compared to an inert wall.
\end{abstract}
\maketitle

\ \\

Understanding how polymer molecules behave near metal surfaces would
greatly enhance our ability to control essential interfacial
properties in a wide variety of problems, including adhesion, wetting
and nano-dewetting, biomolecular recognition, and self-assembly, to
name a few ~\cite{book1996,Theodorou}.  Such an understanding is, however,
difficult to obtain, because widely disparate length scales come into
play ~\cite{megapaper}.  Specifically, atomic-scale energetics
dominate at the interface, where chemi- and physisorption of different
small parts of the polymer molecules may occur. Such details of the
interaction in turn affect the entropically-governed shapes of entire
molecules and thus the bulk properties of the melt close to the surface.  Gaining a better
understanding of the fundamental nature of such an interface thus
requires a multiscale modeling approach.  The purpose of this letter
is to describe a novel multiscale simulation for
molecular-level modeling of polymers near metal surfaces.  Liquid
polycarbonate near a nickel surface is taken as a first example. It
displays specific challenges to the method and at the same time is of
high technical relevance, where
fine control over material properties at the polymer-die interface is
crucial (e.g. production of CD's).

\begin{figure}
\centerline{\psfig{figure=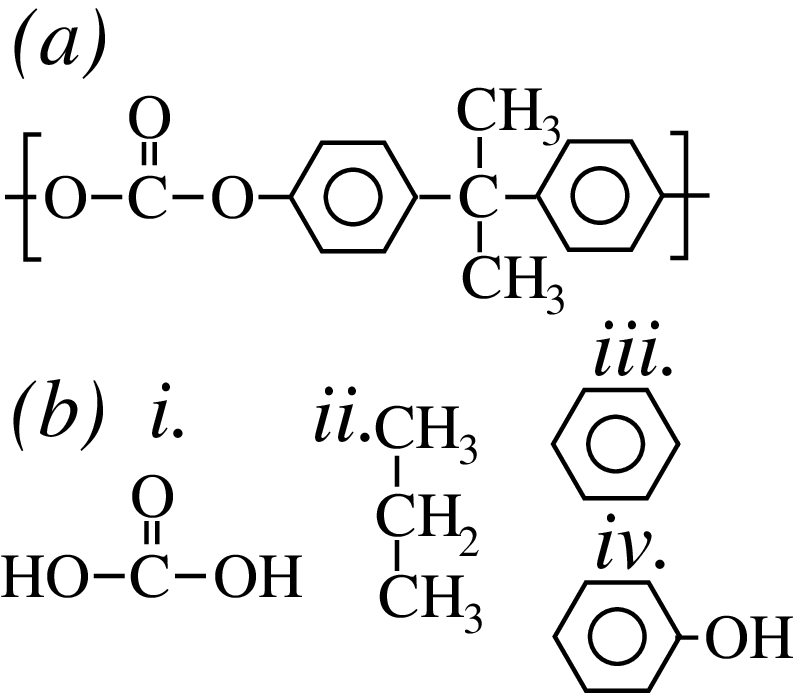}}
\caption{(a) Chemical structure of the repeat unit BPA-PC. 
         (b) Analogous molecules used in the {\it ab initio} studies: 
              (i) carbonic acid, 
             (ii) propane,
            (iii) benzene, and (iv) phenol.}
\label{fig:chain}
\end{figure}

We combine {\it ab initio} calculations for the interaction of
fragments of bisphenol-A-polycarbonate (BPA-PC;
[Fig~\ref{fig:chain}(a)])
and Ni with molecular-level coarse grained simulations of the polymer
melt near a wall. In most polymerization processes for PC, the chain
ends are phenoxy rings.  We find that the adsorption of the phenoxy
chain ends significantly alters the local melt structure.  Such
information is difficult to obtain directly from experiment and bears
specific relevance to both the design of polymer recipes and die
surfaces for optimal processing.  Since the fragments
considered here are not unique to BPA-PC, the results are of general
importance for many organic polymers including biopolymers close to metal surfaces.  

We first describe the ab initio part of our
work. Then we introduce the coarse grained model and its modification
to incorporate specific surface interactions and apply this
to the PC melt near a  Ni surface. 

Our motivation to perform {\it ab initio} calculations is to
ultimately understand how BPA-PC molecules interact with
well-characterized surfaces.  Such calculations are so expensive,
however, that the study of even a single unit of BPA-PC 
near a Ni
surface is not feasible.  Our strategy is to cut the chain into
comonomeric molecules, small enough to study the interaction with the
surface.  We consider three molecules analogous to the comonomeric
subunits of BPA-PC [Fig~\ref{fig:chain}(b)]: carbonic acid (i),
propane (ii)
and benzene (iii) representing carbonate,
isopropylidene, and phenylene, respectively.  Additionally, to
systematically test our choices for the small molecule analogues, we
also consider phenol (iv). Moreover, phenol itself on a metal surface is
important, due to its toxicity, catalytic activity, and widespread
occurrence as a byproduct.  For our purposes here, however, we focus
on how these molecules behave {\em as BPA-PC comonomers}, and examine
the adsorption on a Ni \{111\} surface.  For benzene on
Ni\{111\}, we found an adsorption
energy and geometry in good agreement with Refs.~\cite{Mittendorfer2001,Jenkins1,Jenkins2}. The main results to
emerge from our ab-initio calculations, and incorporated into the
coarse-grained model, are that the carbonic acid and propane molecules
do not stick to the surface.  Benzene, which has a strong
adsorption in isolation, is sterically hindered to adsorb when
incorporated into a BPA-PC chain, due to neighboring carbonate and
isopropylidene groups.  However, phenoxy end groups are not in
this way sterically hindered, and hence may adsorb strongly to the
surface.

We used the plane-wave pseudopotential CPMD code, ~\cite{codecpmd},
implemented with finite-temperature density functional theory
\cite{Alavi1994,AlaviBook1996}. The orbital cut-off was set to 60 Ry.
We used the PBE\cite{Pbe96} generalized gradient approximation (GGA).
The surface is represented by four close-packed layers of Ni \{111\}
(lattice parameter $a_0$ = 3.543\ \AA), with the top two layers
allowed to relax. We used a ($2\times2$) lateral supercell for
carbonic acid and propane adsorption, and a ($3\times3$) cell for
benzene and phenol, employing $4\times4\times1$ and $3\times3\times1$
$k$-point mesh for the smaller and larger cells, respectively.
Several geometry optimizations, starting from plausible structures
compatible with possible orientations of the respective comonomers in
a polymer chain, were performed at each of the four high-symmetry
sites of the \{111\} surface.  The adsorption energy ($E_{\rm ad}$),
Table \ref{table1}, defined as the energy of the 
adsorption system relative to the clean
surface and isolated molecule, characterizes the strength of the
interaction of each submolecule with the surface. 
\begin{table}
\begin{tabular}{|l|c|c|c|c|r|}
\hline
$E_{ad} (eV)$&propane&carbonic acid&benzene&phenol\\
\hline \hline
FCC&0.01&0.01&&0.79\\
HCP&0.01&0.01&&0.84\\
Atop&0.01&0.01&&0.02\\
Bridge&0.01&0.01&1.05&0.91\\
\hline
\end{tabular}
\caption{Adsorption energies at high symmetry sites.}
\label{table1}
\end{table}

For both carbonic acid and propane, $E_{\rm ad}$ turned out to be
rather small, 0.01 eV, which is both negligible compared to the inherent error 
due to the GGA and smaller than the characteristic
thermal energies in typical melt processing of polycarbonate.
The adsorption energies were not significantly changed by going
to a larger cell (3$\times$3), i.e. lower coverage.  
Significant repulsion was experienced by both propane and carbonate
below 3.2\ \AA, regardless of orientation, implying that achieving
distances smaller than this value are highly improbable. 

Benzene ~\cite{nnote}, however, experiences a strong adsorption energy of $E_{\rm
  ad} =$ 1.05 eV at a center-of-mass distance $\approx$ 2~\AA\ from
the surface and in a horizontal orientation, in very good agreement
with ~\cite{Mittendorfer2001}.  Since the carbonate and isopropylidene
moities show strong repulsion for distances shorter than 3.2~\AA,
phenylene is sterically forbidden by its neighbours to approach the
surface to its ideal adsorption distance. We therefore examined the
interaction of benzene with the surface at larger distances, finding
that adsorption is short ranged and decays below 0.03 eV at a distance beyond
3 \AA. Therefore an {\em internal} phenylene comonomer has only a weak
interaction with the surface. For phenol, we calculate an $E_{\rm
  ad}=0.9$ eV, at a distance of 2.0\AA.  Because no steric hindrance
to the horizontal approach of a phenoxy end group to the surface
exists, it is likely that {\em only} the chain ends adsorb strongly to
the surface.  We incorporate the above information into a
coarse-grained model of BPA-PC melts next to a Ni\{111\} wall.  More
details and results of the {\it ab initio} study will be presented in
a forthcoming publication ~\cite{forthcom}.
%\section{Coarse-Grained Polycarbonate Model and Simulation
%  Approach }

\begin{figure}
\centerline{\psfig{figure=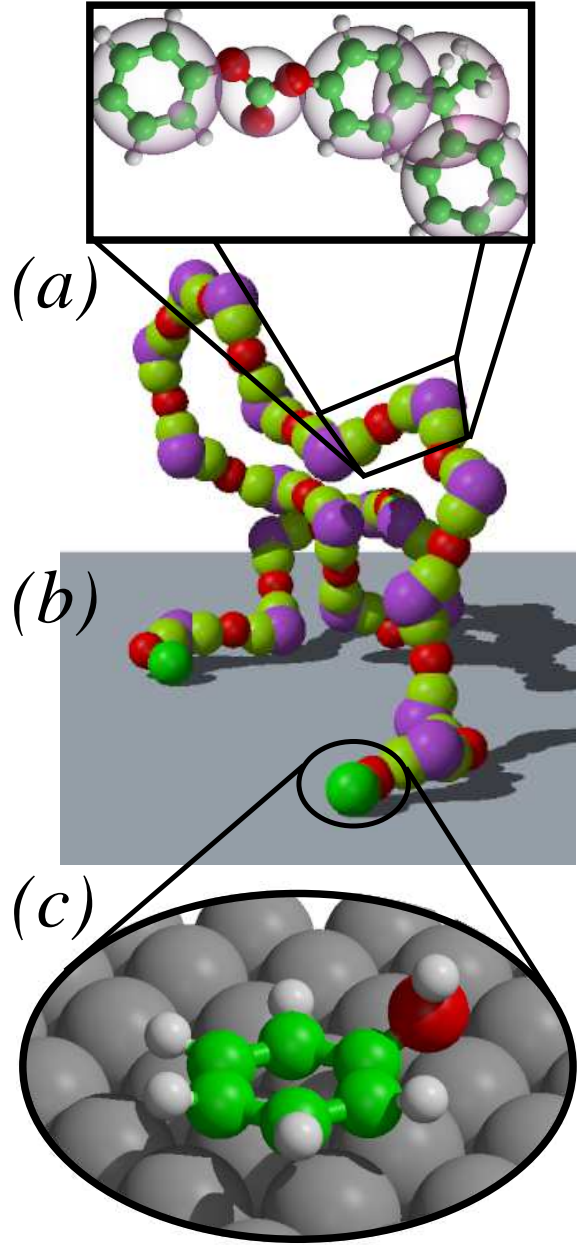,width=75mm}}
\caption{The multiscale model of BPA-PC on nickel.
  (a) The coarse-grained representation of a BPA-PC segment; the
  coarse-grained beads are transparent spheres, superimposed on the
  underlying chemical structure, where the carbon atoms are green, the
  oxygens red, and the hydrogens white; (b) Coarse-grained
  model of an $N$ = 20 BPA-PC molecule, with ends adsorbed on
  a flat surface; configuration from a 160-chain liquid simulation.
  (c) A phenol molecule adsorbed on the bridge site of a (111) nickel
  surface; configuration computed via CPMD simulation.  }
\label{fig:fig2}
\end{figure}

Our polymer model is based on a previously presented coarse-graining
technique~\cite{Tschoep98a}, in the meanwhile improved  ~\cite{Abrams2002inprep}.  The model reduces the number of degrees
of freedom (``particles'') required to faithfully simulate specific
polymer melts at a mesoscopic level to a minimum.  Such a coarsening
is necessary to generate equilibrated samples of long
polymer chain liquids~\cite{Tschoep98a}, whose many different
conformations are governed by the intrachain entropy, $S \propto N$, $N$ 
the number of repeat units.  Briefly, the atomic
structure is mapped onto a bead-spring chain, where each bead, or
mapping point, represents a specific comonomer
[Fig.~\ref{fig:fig2}~(a)], resulting in four beads per BPA-PC repeat
unit.
This requires the enforcement of (1) fixed bead-bead bond lengths along the
backbone; (2) specific bond angle distributions at each bead type; and
(3) torsional states around bead-bead bonds along the backbone. The
bond angle potential around the carbonate-type bead and the torsional
potential at the phenylene connection between carbonate and
isopropylidene beads are obtained from a Boltzmann inversion of
distribution functions of an all-atom simulation of isolated chains at 
the desired temperature, which is 570K in our case.
The average bead diameter and intermolecular length scale, which
reflects the excluded volume of the molecules, fixes the mass and
volume density of the melt~\cite{Tschoep98a}.  Finally, planar
Lennard-Jones 10--4 potentials represent the smooth confining
walls~\cite{Abrams2001b}.  The bead-specific distances at which
monomer subunits experience an average 1 $kT$ repulsion is taken
directly from the {\it ab initio} calculations.  For the phenoxy end
groups and an attractive wall, the 10--4 potential is modified to show
a well depth of 5~$kT$ at a distance of 2.7~\AA.  This distance is
chosen because we represent the phenoxy end group as a sphere 
[Fig.~\ref{fig:fig2}~(b)], and
the 5~$kT$/2.7~\AA\ potential reflects to first order an average over
all near-surface orientations of phenol, including the
horizontally-oriented strongly adsorbed state ~\cite{Abramsnote}.

We simulated melts of 160 chains of 20 repeat units (83 beads per
chain, or a molecular weight of 5292), in a rectilinear box ($L\times
L\times 2L$), by straightforward molecular dynamics.  The runs are
long enough to equilibrate the chain conformations in bulk and near
the surface~\cite{Abrams2001b}.  

%\section{Coarse-Grained Simulation Results}

Because phenoxy chain ends are likely to stick strongly to the
surface, we focus on this effect in the coarse-grained BPA-PC model.  To
do this, we consider the number density of phenoxy end groups,
$\rho_p$, as a function of distance from the walls, $z$. To compare to 
experiments for spatially-resolved
compositional analyses, a more
natural quantity to consider is a cumulative normalized function,
$F(z)$, defined as
\begin{equation}
F(z) \equiv \frac{1}{\rho_{p0}z}\int_0^z \rho_p(z')dz',
\label{eq:F}
\end{equation}
where $\rho_{p0}$ is the bulk density of chain ends.  $F(z)$ is the
ratio of the number of end groups in a layer between 0 and $z$ and the
number expected in such a layer given the overall density of the
system {\it in the absence of walls}. $F(z)$ can be referred to as an
intensity, which is the natural observable of some depth-resolved
experimental techniques, such as angle-resolved
XPS~\cite{Cumpson1995}, none of which, to our knowledge, have yet been
applied to polycarbonate/nickel interfaces.

\begin{figure}
\centerline{\psfig{figure=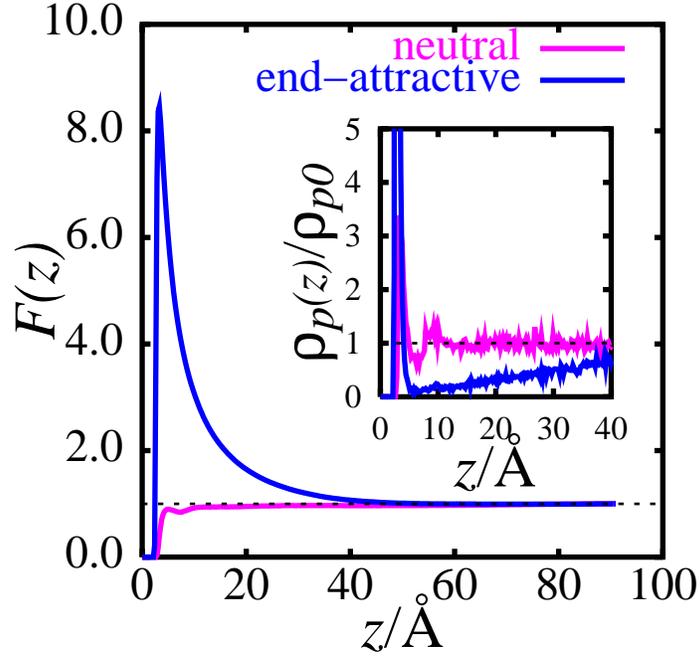}}
\caption{Cumulative normalized density of chain ends, $F(z)$
  (Eq.~\ref{eq:F}), {\it vs} distance from wall, $z$, for a neutral
  wall and a wall displaying strong attraction to chain end beads
  (10-4 Lennard-Jones potential with $\sigma$ = 2.7~\AA\ and
  $\epsilon$ = 5$kT$) for a melt of 160 20-repeat unit BPA-PC chains
  at mass density 1.05 g/cc and temperature $T$ = 570K.  Inset shows
  normalized number density of phenoxy ends, $\rho_p/\rho_{0p}$, {\it
    vs} $z$ for $z < 40$\AA.  For clarity, the $y$ maximum of the
  inset is truncated at 5.0, though $\rho_p(z)/\rho_{0p}$ for the
  attractive ends has a peak-maximum of about 45.0.}
\label{fig:dp}
\end{figure}

In Fig.~\ref{fig:dp}, we show $F(z)$ for the case in which the walls
are completely neutral, and for the case in which the phenoxy end
groups are allowed to adsorb via the energetic interaction discussed
previously.  Also shown in the inset is $\rho_p(z)$ for $z <$ 40 \AA.
For neutral walls the cumulative density quickly rises to the bulk
value and remains constant.  For this case there is a weak
localization of phenoxy end groups at the surface (Fig.~\ref{fig:dp},
inset), which is balanced by depletion on a length scale about one
repeat unit ($\sim$ 8\ \AA).  However, the cumulative normalized density
never exceeds unity.

The situation is dramatically different when the chain ends are
preferentially attracted to the walls.  Here, $F(z)$ displays a
steeply rising overshoot, which shows that the surface has a roughly
eight-times excess of end groups.  Though $F(z)$ decays over long
length scales relative to the attractive potential, 
the chain ends are strongly localized near
the surface (inset Fig.~\ref{fig:dp}).  Also there is a large
depletion zone of chain ends in the liquid adjacent to this strongly
adsorbed layer.  As $z$ increases, the end group intensity decays to
the bulk value at about 40 \AA\ from the wall, which is larger than
the mean gyration radius of about 30 \AA. Viscoelastic
consequences will be the subject of future work.

These results are intriguing in light of previous
simulations~\cite{Bitsanis90} of simple bead-spring chains, which
showed that neither neutral or attractive walls produced chain end
depletion layers that reflect the size of the chains.  Our results
indicate that walls that adsorb only chain ends do indeed show 
such a region in the melt.  This highlights the importance of chain
end mobility  (and small molecule impurities) in the overall
composition of a thin adsorbed polymer film.

In contrast, more recent mean-field calculations~\cite{Wu95}
contend that natural enhancement of chain ends at neutral surfaces
results in near-surface excess ends with an adjacent depletion zone
whose depth reflects the size of the chains.  Our results indicate
that this is not a generic property of polymers.  Instead, we find
that this is only observed with an explicit end-surface attraction.
When no such attraction is given, perturbations in end-group
distribution in the liquid die out much less than a depth of $R_g$
from the wall.  The discrepancy arises from the interplay between the
intrachain statistics and the packing of the irregularly-shaped
monomers near the surface. 

%\section{Conclusion}

By employing a multiscale approach, we showed how the interplay of
entropic and energetic contributions can alter the structure of a
polymer melt near a metal surface.  Specifically, we predict that
polycarbonate chain ends adsorb strongly to a nickel surface next to a
polycarbonate liquid.  This suggests modifying chain ends provides a
reasonably sensitive way to control interfacial behavior of
polycarbonate in the manufacture of optical data storage devices, and
other modern nanostructured surfaces. The technique of marrying {\it ab
initio} calculations involving small molecule analogues of polymeric
comonomers with chemically-specific coarse grained models opens up the
possibilty to study a wide array of biomolecule/metal surface
interactions.  Thus this study sheds a new light on the technically
important field of polymer/metal surface interactions and highlights
the important need for experimental contributions.

{\it Acknowledgements}: We are grateful to M. Parrinello for 
suggestions and for providing us the CPMD code, and to F.
M\"uller-Plathe, G. Wegner, and B. Jones for comments and
suggestions. The work was supported by the Bundesministerium f\"ur
Bildung and Forschung, BMBF, grant No.03 N 6015.  Images were rendered using the Raster3D
package.~\cite{Merritt97}

%\bibliographystyle{unsrtnat}

%\bibliography{references,specifics}

\end{document}